\documentstyle[aps,prl,floats,graphicx,epsfig]{revtex}
\begin{document}
\draft

\twocolumn[\hsize\textwidth\columnwidth\hsize\csname @twocolumnfalse\endcsname

\title{Effective spacetime and Hawking radiation from moving domain wall in
thin film of $^3$He-A.}

\author{ T.A. Jacobson$^{1}$ and G.E. Volovik$^{2,3}$}
\address{
$^{1}$Department of Physics, University of
Maryland, College Park, MD 20742-4111, USA\\
$^{2}$Low Temperature Laboratory, Helsinki University of
Technology,
P.O.Box 2200, FIN-02015 HUT, Finland\\
$^{3}$L.D. Landau Institute for
Theoretical Physics,  Kosygin Str. 2, 117940 Moscow, Russia }

\date{\today}
\maketitle

\begin{abstract}
An event horizon for ``relativistic" fermionic quasiparticles can
be constructed in a thin film of superfluid $^3$He-A.  The quasiparticles
see an effective ``gravitational" field which is induced
by a topological soliton of the order parameter.
Within the soliton the "speed of light" crosses zero and changes sign.
When the soliton moves, two planar
event horizons (black hole and white hole) appear, with a curvature
singularity between them. Aside from the singularity, the effective
spacetime is incomplete at future and past boundaries, but the
quasiparticles cannot escape there because the nonrelativistic
corrections become important as the blueshift grows, yielding
``superluminal" trajectories. The question of Hawking radiation
from the moving soliton is discussed but not resolved.
\end{abstract}

\

\pacs{PACS numbers: 04.70.Dy, 67.57.-z, 67.57.Fg    }

\

] \narrowtext

{\it Introduction}.
Condensed matter systems can serve as a useful model to study
problems related to black-hole event
horizons \cite{UnruhSonic,Visser1997}. Recently we found that
  moving textures in a quantum fluid --  superfluid $^3$He-A  --
provide for us the possibility to study quantum properties
of the event horizon,
including Hawking radiation \cite{JacobsonVolovik}. However in that
example the Hawking radiation was essentially masked by Schwinger pair
creation outside the horizon, which appeared to be the main mechanism of
quantum dissipation at zero temperature.  Here we discuss another texture,
where now the Hawking radiation may dominate. This texture is a soliton,
which is topologically stable
in a thin film of superfluid $^3$He-A. Our work is
partially motivated by recent success in experimental study of thin superfluid
$^3$He films, where the density of superfluid component was measured using the
third sound technique \cite{Packard}.

{\it Order parameter and quasiparticle spectrum.}
The orbital part of the order parameter of the A-phase state in thin films
is characterized by a complex vector which is parallel to the
plane of the film:
\begin{equation}
\vec \Psi = {\bf e}_1 + i{\bf e}_2 ~,~{\bf e}_1\perp {\hat{\bf z}}~,~{\bf
e}_2\perp {\hat{\bf z}}~,
\label{OrderParameterAphase}
\end{equation}
where the axis $z$ is along the normal to the film.
This vector characterizes the Bogoliubov-Nambu Hamiltonian for the
fermionic excitations in the $^3$He-A vacuum:
\begin{equation}
 {\cal H}= v_F(p-p_F)~\check\tau _3 + {\bf e}_1 \cdot   {\bf p}~\check \tau_1
- {\bf e}_2 \cdot   {\bf p} ~\check \tau_2   ~ .
\label{FermionHamiltonian}
\end{equation}
where  $\check \tau_a$ are the Pauli matrices in the Bogoliubov-Nambu
particle-hole space, and we neglected the conventional spin degrees of freedom.
The square of the Hamiltonian matrix
${\cal H}_{\rm A}^2=E^2({\bf p})$ gives the square of the quasiparticle
energy spectrum:
\begin{equation}
E^2({\bf p})=  v_F^2(p-p_F)^2 +({\bf e}_1 \cdot   {\bf p})^2 +
({\bf e}_2 \cdot   {\bf p})^2~~.
\label{ESquare}
\end{equation}
The simplest realization of the equilibrium (vacuum) state of $^3$He-A is
${\bf e}_1^{(eq)}=c_\perp {\hat{\bf x}}$ and  ${\bf
e}_2^{(eq)}=c_\perp {\hat{\bf y}}$, where the parameter $c_\perp \sim
3~$cm/sec at zero pressure. All the other degenerate states can be obtained by
the symmetry operations: continuous rotations about the axis $z$ and discrete
$\pi$-rotation about a
perpendicular axis. The  vacuum manifold consists of two
disconnected pieces which can be transformed to each other only by the
discrete transformation. This results in the topological solitons - domain
walls.

{\it Domain wall}
If the domain wall is parallel to the plane $y,z$  the order parameter has the
following form
\begin{equation}
{\bf e}_1(x)=c^x(x) {\hat{\bf x}} ~,~{\bf e}_2(x)=c^y(x) {\hat{\bf
y}} ~~,
\label{OrderParameterGeneral}
\end{equation}
Across the soliton  either the function $c^x(x)$ or
the function $c^y(x)$ changes sign: both cases correspond to the same class of
topologically stable soliton and can transform to each other. The horizon
appears in the former case, and for simplicity we choose the following Ansatz
for such a soliton
\begin{equation}
 c^y(x)= c_\perp~,~c^x(x)= - c_\perp\tanh {x\over d}  ~.
\label{OrderParameterAnzats}
\end{equation}
It is close to the solution for the domain wall obtained in
Ref.\cite{SalomaaVolovik1989} (see Fig. 3b
of Ref.\cite{SalomaaVolovik1989}). Here the thickness of the domain wall $d
\sim
1000\AA$ and is of order the thickness of the film \cite{Packard}.

{\it ``Relativistic" spectrum.}
We are interested in the low-energy quasiparticles, which are
concentrated in the vicinity of momenta ${\bf p}=\pm p_F {\hat{\bf z}}$.
Close to these two points the quasiparticle energy spectrum becomes:
\begin{equation}
E^2({\bf p})=c_\parallel^2(p_z\mp p_F)^2 + (c^x (x)p_x)^2 + c_\perp^2 p_y^2
~,~c_\parallel=v_F~,
\label{RelE}
\end{equation}
up to terms of order $p_\perp^4/m_*^2$, where $m_*=p_F/v_F$.
After shifting the momentum $p_z$ (\ref{RelE})  can be rewritten   in the
Lorentzian form
\begin{equation}
g^{\mu\nu} p_\mu p_\nu  =0 ~~.
\label{RelEquation}
\end{equation}
Here $p_\mu=(-E,p_i)$ is the four momentum, and the nonzero elements
of the inverse metric are given by
\begin{equation}
g^{00}=1~,~g^{zz}= -c_\parallel^2~,~g^{yy}=-c_\perp^2 ~,~g^{xx}=- (c^x
(x))^2~~.
\label{metric}
\end{equation}
Thus in this domain wall the speed $c^x(x)$ of light,  propagating across the
wall, changes sign. This corresponds to the change of the sign of one of
the vectors,
${\bf e}_1$, of the effective vierbein in Eq.(\ref{OrderParameterGeneral}).

{\it Effective space-time induced by stationary soliton.}
The line element corresponding to the
inverse metric in Eq.(\ref{metric}) is
$(ds^2)_{3+1}= (ds^2)_{1+1}  -c_\perp^{-2}dy^2 - c_\parallel^{-2} dz^2$, with
\begin{equation}
(ds^2)_{1+1}=dt^2 - \bigl(c^x(x)\bigr)^{-2}\, dx^2,
\label{flat}
\end{equation}
We emphasize
that the coordinates $t,x,y,z$ are the coordinates of the background
Galilean system, while the interval $ds$ describes the effective Lorentzian
spacetime viewed by the low-energy quasiparticles.

The line element (\ref{flat}) represents a {\it flat} effective spacetime
for any
function $c^x(x)$. The singularity at $x=0$, where $g^{xx}=0$, can be
removed by
a coordinate transformation. In terms of a
new coordinate $\zeta=\int dx/c^x(x)$ the line element takes the standard form
$dt^2 - d\zeta^2$. With  $c^x(x)$ given by (\ref{OrderParameterAnzats})
$\zeta$ diverges logarithmically as $x$ approaches zero.
Thus (\ref{flat}) is actually {\it two copies} of flat
spacetime glued together along the line $x=0$ where $c^x(x)$ vanishes.
This line is an infinite proper
distance away along any spacelike or timelike geodesic.
The two spacetimes are disconnected in the relativistic
approximation, however this approximation breaks down
near $x=0$ and the two halves actually communicate.
One must also keep in mind that invariance under
general coordinate transformations holds
only for the physics of ``relativistic" low-energy quasiparticles,
but not for the background
superfluid and high-energy ``nonrelativistic" excitations. The singularity at
$x=0$ is thus physical, but is unobservable by the low-energy quasiparticles
since the Ricci scalar is zero everywhere.

{\it Effective space-time in moving domain wall.}
Let the soliton move
relative to the
superfluid:   ${\bf v}\equiv {\bf v}_w - {\bf v}_s = v \hat x$, where ${\bf
v}_w$ and $ {\bf v}_s$ are the velocities of the domain wall and superfluid
condensate respectively. The energy spectrum of the quasiparticles is well
defined in the soliton frame where the order parameter is again stationary; it
is Doppler shifted:
\begin{equation}
E({\bf p})=\pm\sqrt{ c_\parallel^2 p^2_z +c_\perp^2 p_y^2 + (c^x(x) p_x)^2}
-p_x
v~.
\label{EDopplerShifted}
\end{equation}
This leads to the following modification of the 1+1 ($t,x$) metric elements:
\begin{equation}
g^{xx}=v^2-(c^x(x))^2~,~g^{00}= 1~,~g^{0x}=-v~.
\label{Metric}
\end{equation}
 Here now $x$ is the ``comoving" coordinate,
at rest with respect to the soliton.
The line element in this 1+1 effective metric takes the form:
\begin{equation}
(ds^2)_{1+1} =dt^2 -  \bigl(c^x(x)\bigr)^{-2}\, (dx+vdt)^2~,
\label{moving}
\end{equation}
which also follows directly from the Galilean
transformation to the moving frame.

With $c_\perp >v> 0$ the effective spacetime geometry is
no longer flat but rather that of a black hole/white hole
pair (Fig.\ref{soliton4}).  The black hole and white hole horizons are located
where
$|c^x(x)|=v$, at positive and negative $x$ respectively:
\begin{equation}
\tanh  {x_h\over d} = \pm {v\over c_\perp} ~.
 \label{Horizon}
\end{equation}
The positions of the horizons coincide with the positions of the ergoplanes,
since the metric element
\begin{equation}
g_{00}= 1- {v^2\over (c^x(x))^2}   ~.
\label{g00}
\end{equation}
crosses zero at the same points as $g^{xx}$.

\begin{figure}[!!!t]
\begin{center}
\leavevmode
\epsfig{file=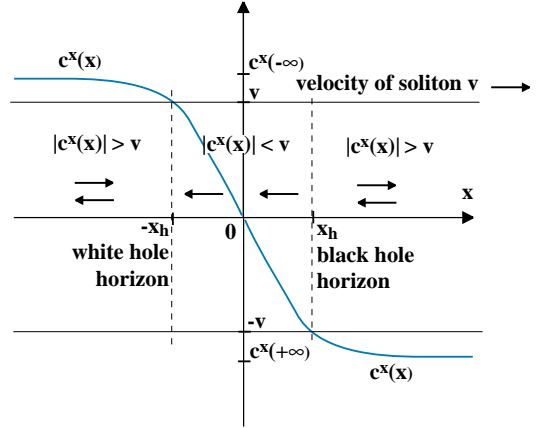,width=0.8\linewidth}
\caption[soliton4]
    {The ``speed of light" in the $x$-direction in the soliton frame,
$c^x(x)$, for the singular soliton in thin films of $^3$He-A. The speed of
light
crosses zero at $x=0$. For the moving soliton the  black and white hole pair
appears for any velocity $v$ below $|c^x(\infty)|$. }
\label{soliton4}
\end{center}
\end{figure}

It follows from Eq.(\ref{Horizon}) that if $v$ approaches the asymptotic value
$c_\perp$ of the speed of light, i.e. $c_\perp - v
\ll c_\perp$, the positions of the horizons move far away from the  $x=0$ line:
$|x_h| \gg d $. At these positions the gradients of the order parameter are
small, so that the quasiclassical spectrum in Eq.(\ref{EDopplerShifted}) and
thus the description in terms of the effective metric can well be applied near
the horizons, even if the thickness of the soliton $d$ is of order $\xi$,
the coherence length.

The line
$x=0$ is now at {\it finite} proper time or distance along some
geodesics (although still at infinite proper distance along a
$t=$ constant line).
For example, $x=-vt$ (a point at rest with respect to the
superfluid) is a geodesic along which $t$
is the proper time, which is clearly finite as $x=0$ is
approached.
Moreover, $x=0$ is now {\it spacelike}, and the curvature diverges
there. It is therefore akin to the singularity at
$r=0$ inside a spherically symmetric neutral black hole.

The Ricci scalar for the line element (\ref{moving})
is (we removed index $x$ in $c^x$)
\begin{equation}
R=2{v^2\over c^2}\bigl(cc''-2(c')^2\bigr)=
-{4v^2\over d^2} \bigl({c_\perp^2\over c ^2}-{c^2\over c_\perp ^2}\bigr)~.
\label{Curvature}
\end{equation}
As $x\rightarrow0$ this diverges like $-(2v/x)^2$
for any nonzero $v$. For $v=0$, the spacetime is flat,
as noted above.
At the horizon $R= -(2c_\perp/d)^2\bigl(1-(v/c_\perp)^4\bigr)$.
Note that as $v\rightarrow c_\perp$ the curvature at the horizon
goes to zero.

\begin{figure}[!!!t]
\begin{center}
\leavevmode
\epsfig{file=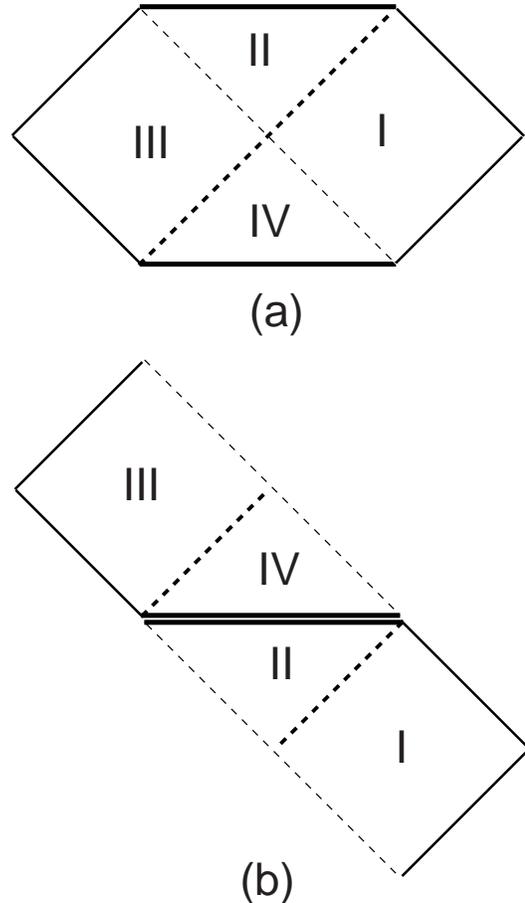,width=0.8\linewidth}
\caption[Penrose]
    {Penrose diagrams. The horizontal thick solid lines represent curvature
singularities. The solid lines at 45$^o$ are at lightlike
infinity.
Fig. (\ref{Penrose}a) is the diagram for a spherically symmetric neutral
(Schwarzschild) black hole. The thick dashed
line is the black hole horizon and the thin dased line is the white hole
horizon. Fig. (\ref{Penrose}b) is the Penrose diagram for the
moving thin film soliton, which can be constructed by cutting
Fig. (\ref{Penrose}a) along the thin dashed line and glueing the
two pieces together along the singularity. Here the thick dashed lines are the
black and white hole horizons at $x=\pm x_h$, while the thin dashed
lines are boundaries where the effective relativistic spacetime is
incomplete. However, the diagram represents all of physical space and
time (see text).}
\label{Penrose}
\end{center}
\end{figure}

The positive $x$ piece of (\ref{moving}) has the causal structure
of regions I and II of the Penrose diagram (Fig. \ref{Penrose}(a))
for the radius-time section of a Schwarzschild black hole,
while the negative $x$ piece has the structure of regions III and IV.
These two pieces fit together as shown in  Fig. \ref{Penrose}(b).

The causal diagram reveals that the physical ranges of the
coordinates $t$ and $x$ do not cover a complete
manifold in the sense of the line element (\ref{moving}).
Geodesics of finite length can run off the thin dashed line
boundary at the lower edge of region
I or the upper edge of region III in \ref{Penrose}(b). This at
first appears paradoxical: how could a quasiparticle escape
from physical space and time in a finite ``proper" time
(or affine parameter along a lightlike geodesic)?
The answer is that the energy in the superfluid frame
would diverge at the edge where $t$ goes to infinity
(due to the ``gravitational blueshift"), but before it actually
diverges the higher
order terms in the dispersion relation (\ref{ESquare}) become
important, and the quasiparticle is deflected from the
trajectory governed by the metric (\ref{moving}). For example,
if we follow an outgoing q.p. backwards in time towards the horizon,
relativistically it would run off region I into
region IV in Fig. \ref{Penrose}(a). In fact, however, as it
gets close to the horizon, its momentum grows
until (\ref{EDopplerShifted}) no longer holds.
The higher order term
$p_\perp^4 (v_F/2p_F)^2$ in the dispersion relation (\ref{ESquare})
gives the q.p. a ``superluminal" group velocity $v_F p_\perp/p_F>c_\perp$
at large $p_\perp$, so it crosses the
horizon backwards in time and runs into the singularity. Whether it survives
this encounter with the singularity we do not yet know.

{\it Quantum dissipation by Hawking radiation.}
In the presence of a horizon the vacuum becomes ill-defined,  which leads to
dissipation during the motion of the soliton.  One of the mechanisms of
dissipation and friction may
be the Hawking black-body  radiation from the horizon
\cite{Hawking}, with temperature determined by
the ``surface gravity":
\begin{equation}
T_H={\hbar\over 2\pi k_B} \kappa~,~\kappa= \left({dc^x\over dx}\right)_h
~~.
\label{HawkingT}
\end{equation}
In our case of Eqs.(\ref{OrderParameterAnzats},\ref{moving}) the Hawking
temperature depends on the velocity $v$:
\begin{equation}
T_H(v)=T_H(0) \left( 1- {v^2\over c_\perp^2}\right)~~~,~~~T_H(0)={\hbar
c_\perp
\over 2\pi k_B d}~.
\label{T(v)}
\end{equation}
Typically $T_H(0)\sim 1~\mu K$, however we must choose $v$ close to $c_\perp$
to make the relativistic approximation  more  reliable. This Hawking
radiation could in prinicple
be detected by quasiparticle detectors. Also it leads to
energy dissipation and thus to deceleration of the moving domain wall even if
the real temperature $T=0$. Due to deceleration caused by Hawking
radiation, the
Hawking temperature increases with time. The distance between horizons, $2x_h$,
decreases until the complete stop of the domain wall when the two horizons
merge (actually, when the distance between them becomes comparable to the
``Planck length" $\xi$).  The  Hawking temperature approaches its asymptotic
value
$T_H(v=0)$ in Eq.(\ref{T(v)}); but when the horizons merge, the Hawking
radiation disappears: there is no more ergoregion with
negative energy states, so that the stationary domain wall is nondissipative,
as it should be.

At the moment, however, it is not very clear whether and how
the Hawking radiation occurs in this
system.
There are several open issues: (i) The particle conservation
law may prevent the occupation of the negative energy states behind the
horizon.
(ii) Even if these states can be occupied they may saturate, since the
quasiparticles are fermions, thus
cutting off further Hawking radiation.  (iii)  The
appropriate boundary condition for the Hawking effect may not hold:
the outgoing high
frequency modes near the horizon should be in their ground state. These modes
come from the singularity, since they propagate "superluminally", so the
physical question is whether, as the singularity moves through
the superfluid, it excites these modes or not.
(iv) Even if this boundary condition holds initially, the mechanism
discussed in \cite{CorJac}
of runaway damping of Hawking radiation for a superluminal fermionic
field bewteen a pair of horizons may occur. (v) Although the Hawking
temperature (\ref{T(v)}) can be arbitrarily low for $v$ near $c_\perp$,
this is
the temperature in the frame of the texture. Perhaps more relevant to
the validity ofthe relativistic approximation is the temperature in the
frame of the superfluid, which is given by $T_{sf}=T_H(0)(1+v/c_\perp)$.
This is never lower than $T_H(0)\sim \hbar c_\perp/d\sim m_*c_\perp^2$
if $d\sim\xi$,
which is just high enough for the nonrelativistic corrections to be
important near the peak of the thermal spectrum. Thus, unless the
width of the soliton $d$ can be arranged to be much greater than the
coherence length $\xi$, only the low energy tail of the radiation
will be immune to nonrelativistic corrections.
We leave these problems for further investigation.

We expect that as $v\rightarrow0$, the entropy of the domain wall approaches a
finite value, which corresponds to  one degree of quasiparticle freedom per
Planck area. This is similar to the  Bekenstein entropy \cite{Bekenstein}, but
it comes from the ``nonrelativistic'' physics at the ``trans-Planckian'' scale.
It results from the fermion zero modes: bound states at the domain wall with
exactly zero energy. Such bound states, dictated by topology of the texture,
are now intensively studied in high-temperature superconductors and other
unconventional superconductors/superfluids (see references in \cite{Calderon}
and \cite{1/2vortex}). When $v\ne0$, there must be another contribution
to the entropy, which can be obtained by integrating
$dS = dE/T$.

This work was supported by the Russian Foundation for Fundamental Research
grant No. 96-02-16072, by the Intas grant 96-0610, by the European Science
Foundation, and by National Science Foundation
grants PHY94-13253 and PHY98-00967.

\end{document}